\documentclass[a4paper,twocolumn,11pt]{quantumarticle}
\pdfoutput=1
\usepackage[utf8]{inputenc}
\usepackage[english]{babel}
\usepackage[T1]{fontenc}
\usepackage{amsmath}
\usepackage{hyperref}
\usepackage{amsmath}  % For \left and \right brackets
\usepackage{amssymb}  % Provides \mathbb
\usepackage{float}

\usepackage{tikz}
\usepackage{lipsum}

\begin{document}

\title{Adaptive Entanglement Routing with Deep Q-Networks in Quantum Networks}

\author{Lamarana Jallow}
\affiliation{Department of Computer Science, Comsats University Islamabad,  Islamabad, Pakistan}
\affiliation{Department of Computer Science, University of The Gambia, Faraba, The Gambia}

\author{Majid Iqbal Khan}

\affiliation{Department of Computer Science, Comsats University Islamabad,  Islamabad, Pakistan}

\maketitle

\begin{abstract}
The quantum internet holds transformative potential for global communication by harnessing the principles of quantum information processing. Despite significant advancements in quantum communication technologies, the efficient distribution of critical resources, such as qubits, remains a persistent and unresolved challenge. Conventional approaches often fall short of achieving optimal resource allocation, underscoring the necessity for more effective solutions. This study proposes a novel reinforcement learning-based adaptive entanglement routing framework designed to enable resource allocation tailored to the specific demands of quantum applications. The introduced QuDQN model utilizes reinforcement learning to optimize the management of quantum networks, allocate resources efficiently, and enhance entanglement routing. The model integrates key considerations, including fidelity requirements, network topology, qubit capacity, and request demands. In comparison to existing algorithms, QuDQN demonstrates superior performance, achieving up to 16.56\% higher throughput, resolving 38.54\% more requests under high-demand conditions, and reducing qubit usage by up to 38.34\%. These results highlight the model’s enhanced efficiency and its potential to enable scalable and robust quantum communication infrastructure.
\end{abstract}

The quantum internet has the potential to revolutionize global communication by leveraging quantum information processing to enable faster, more secure, and highly efficient communication systems \cite{bib1}. Projects like the DARPA quantum network \cite{bib3}, Tokyo QKD network \cite{bib5}, SECOQC Vienna QKD network \cite{bib4}, and China’s satellite quantum network \cite{bib6}, along with applications such as secure key sharing, sensing, and synchronisation \cite{bib7}, \cite{bib8}, \cite{bib9}, \cite{bib10}, show how far this technology has come. As quantum technology moves closer to real-world use, the quantum internet could work alongside or even replace parts of the current internet, becoming a key component of future internet architecture. \cite{bib11}.

Communication protocols are key to making these technologies work, and quantum entanglement plays a crucial role in sending qubits between distant nodes \cite{bib12}. Quantum nodes, connected by optical links, are needed to create, store, and process quantum information \cite{bib13}. Quantum entanglement not only makes data transmission secure but also helps establish routing paths for sending entanglement over long distances. To extend entanglement over longer distances, nearby nodes create entangled pairs using a swapping process, with help from quantum repeaters and classical communication. However, because quantum repeaters and systems are complex, setting up efficient ways to distribute entanglement is challenging \cite{bib14}. 

Quantum communication applications have different needs for entanglement, much like how classical networks handle various types of services. Some applications focus on secure communication, while others need fast task completion to reduce the effects of decoherence in quantum memories, ensuring timely results \cite{bib2}. Quantum links, often using optical technology, have limitations that make it harder to distribute entanglement over short distances \cite{bib16}. For large-scale quantum networks, a multi-hop structure will be necessary, where quantum routing creates paths between source and destination nodes. This involves assigning the right entanglements based on qubit connections \cite{bib17}. 

Reinforcement Learning (RL) and machine learning are increasingly being used to automate tasks in quantum information science, as shown in several studies \cite{bib34}, \cite{bib35}, \cite{bib36}. For instance, machine learning has shown great potential in helping design and optimize quantum experiments \cite{bib33}. While deep learning has made a big impact in many areas, its use for routing in quantum networks is still limited. This research uses reinforcement learning to adapt to the changing nature of quantum network resources, efficiently allocating them for communication requests. The aim is to find the best routing paths and create entanglement quickly while meeting the required fidelity levels.

\textbf{\textit{Problem Formulation:}}
Entanglement routing in quantum networks has predominantly focused on evaluating conventional routing algorithms, such as Dijkstra's, multipath routing, and greedy routing, within specific network configurations, as evidenced by recent studies \cite{bib29}, \cite{bib30}, \cite{bib31}, \cite{bib32}. However, these investigations often rely on simplified topologies, such as ring structures, and base their assumptions on these idealized models. Such simplifications may overlook the practical complexities inherent in real-world quantum networks, which typically exhibit dynamic and intricate architectures. Unlike fixed topologies, quantum networks must adapt to the placement of end hosts in specific locations dictated by the requirements of quantum applications. This necessity for flexibility introduces significant challenges in routing, as real-world quantum networks are neither simple nor static.

A critical challenge in advancing routing techniques is the fragmented nature of research on entanglement routing and the lack of comprehensive evaluations tailored to quantum networks. Existing studies frequently concentrate on generic network structures rather than addressing the unique demands of quantum applications. For instance, studies such as \cite{bib10}, \cite{bib39}, and \cite{bib48} highlight the importance of qubit availability at quantum nodes for entanglement routing. However, this approach often results in low fidelity and insufficient entanglement generation rates due to frequent failures. These failures lead to delays, as requests must await the next available cycle, making it difficult to satisfy the stringent requirements of quantum applications. While these studies propose basic algorithms for qubit allocation, they are often inadequate for managing large-scale networks and handling multiple simultaneous requests. Furthermore, they lack the adaptability needed to address the dynamic nature of quantum networks, significantly limiting their overall effectiveness.

The rest of the paper is organized as follows: Section 2 discusses the basics of quantum communication and reinforcement learning. Section 3 discusses related work. Section 4 presents the proposed model and Section 5 focuses on performance analysis, and lastly, Section 6 concludes the whole discussion.

\section{Quantum Basics and Reinforcement Learning}\label{sec3}
In this section, we look at the core principles of quantum information processing and communication, and Reinforcement Learning. This gives a thorough understanding of the key concepts and underlying principles. 

\subsection{Quantum Information Processing  and Communication}
Quantum information processing forms the foundation of the quantum internet, promising to revolutionize global communication \cite{bib1}. It relies on the principles of quantum mechanics to manipulate and transmit information encoded in quantum bits (qubits). Unlike classical bits, which can only be in states 0 or 1, qubits can exist in a superposition of these states \cite{bib11}. The state of a qubit, \(\vert \psi \rangle\), is expressed as a linear combination of the basis states \(\vert 0 \rangle\) and \(\vert 1 \rangle\):

\begin{equation}
\vert \psi \rangle = \alpha \vert 0 \rangle + \beta \vert 1 \rangle  \label{eq1}
\end{equation}

where \(\alpha\) and \(\beta\) are complex amplitudes satisfying  \(|\alpha|^2 + |\beta|^2 = 1\).

Quantum entanglement occurs when the states of two or more particles become correlated, such that the state of one particle cannot be described independently of the others \cite{bib11}. This enables instantaneous information transfer over distances. For two qubits, the entangled state is represented as:
\begin{equation}
 \frac{1}{\sqrt{2}} \left( \vert 00 \rangle + \vert 11 \rangle \right)  
\end{equation}
Entanglement Swapping is a key operation in quantum communication, allowing entangled states to extend across multiple quantum nodes. This process, facilitated by quantum repeaters, involves two pairs of entangled qubits (e.g., AB and CD) to create a new entangled state between particles A and D, which have never directly interacted. The joint state of the four particles is:

\begin{equation}
 \vert \Psi^{AB} \rangle \otimes \vert \Psi^{CD} \rangle  
\end{equation}
Now, if particles B and C are measured in a Bell basis the resulting outcome is one of the four Bell states:
\begin{equation*}
 \frac{1}{\sqrt{2}}(\vert 00 \rangle + \vert 11 \rangle), \quad \frac{1}{\sqrt{2}}(\vert 00 \rangle - \vert 11 \rangle),
 \end{equation*}
 \begin{equation}
\quad \frac{1}{\sqrt{2}}(\vert 01 \rangle + \vert 10 \rangle), \quad \frac{1}{\sqrt{2}}(\vert 01 \rangle - \vert 10 \rangle)
\end{equation}
If the first outcome, \(\frac{1}{\sqrt{2}}(\vert 00 \rangle + \vert 11 \rangle)\) is obtained, particles A and D become entangled. Their joint state is:

\begin{equation}
\frac{1}{\sqrt{2}}(\vert 00 \rangle + \vert 11 \rangle)
\end{equation}

This Bell state measurement enables entanglement swapping, transferring entanglement from the original pairs (AB and CD) to the new pair (AD). This process is crucial for quantum repeaters, which generate entanglement over long distances in quantum networks \cite{bib1}.

In a quantum repeater network, nodes can generate and store entangled pairs of qubits \cite{bib14}. Consider a network with nodes labeled \(N_1, N_2, \ldots, N_N\). When qubit pairs are entangled, they form the basis for entanglement connections between nodes. The entangled pair \(\vert \Psi \rangle\) is represented as:
\begin{equation}
\vert \Psi \rangle = \alpha \vert 0 \rangle + \beta \vert 1 \rangle 
\end{equation}

Quantum routing is essential for quantum networks, enabling the establishment of entanglement paths between source and destination nodes. Unlike classical networks, quantum networks leverage entanglement, a feature absent in classical systems \cite{bib23}. In a quantum network with \( N \) nodes, each node \( N_i \) generates entangled pairs \( \vert \Psi_i \rangle \). The goal of quantum routing is to establish an entanglement connection between a source node \( N_s \) and a destination node \( N_d \). This is achieved by selecting an appropriate entangled pair \( \vert \Psi_i \rangle \) from \( N_s \), sending one qubit to \( N_d \) while retaining the other at \( N_s \). The entangled pair is represented as:
\begin{equation}
\vert \Psi_i \rangle = \alpha_i \vert 0 \rangle + \beta_i \vert 1 \rangle.
\end{equation}

The routing decision involves selecting the optimal entangled pair \( \vert \Psi_i \rangle \) based on parameters such as fidelity and resource availability. Algorithms for this purpose optimize qubit allocation to meet the demands of specific quantum communication tasks.

\subsection{Reinforcement Learning }
Reinforcement learning is a deep learning approach used for adaptive decision-making, making it well-suited for tackling challenges in dynamic and complex environments. Among RL techniques, Deep Q-Networks (DQN) stand out for their ability to handle large state spaces and provide optimal solutions in real-time scenarios. DQN trains an agent to take actions that maximize cumulative rewards over time, making it a powerful tool for improving the performance of quantum networks. These networks are expected to form the foundation of next-generation communication and information processing systems \cite{bib20}. In quantum networks, RL can be applied to dynamically allocate resources, such as qubits, and optimize entanglement routing.

A RL problem is described by a tuple \((\mathcal{S}, \mathcal{A}, \mathcal{P}, \mathcal{R})\), where:
- \(\mathcal{S}\) refers to the state space—that is, all possible states of the quantum network.
- \(\mathcal{A}\) represents the action space, which encapsulates the set of all feasible actions given a certain state.
- \(\mathcal{P}\) describes the transition probability function, detailing the quantum network dynamics.
\(\ (\mathcal{R}\) ) be the reward function, which  assigns the desirability of state-action pairs.

The goal of RL is to find a strategy \(\pi: \mathcal{S} \rightarrow \mathcal{A}\) that maximises the predicted cumulative reward over a long time.

\subsection{Deep Q-Networks}

The DQN method combines RL with deep neural networks. It approximates the Q-function to predict the expected cumulative reward of choosing an action in a given state \cite{bib21}. The Q-function is defined as
\begin{equation}
Q(s,a) = \mathbb{E}\left[R_{t} + \gamma \max_{a'} Q(s',a') \right]
\end{equation}
Where $(s)$ represents the current state, $R_{t}$ gives the immediate reward, $(a)$ is the action taken, $s'$ is the next state, $\gamma$ is the discount factor and $a'$ gives the next action

The DQN algorithm iteratively updates its Q-network to minimize the temporal difference error, defined as 
\begin{equation}
\delta = Q(s,a) - (R_{t} + \gamma \max_{a'} Q(s',a'))
\end{equation}
DQN uses a deep neural network to approximate the action-value function. The network receives the state $s$ as input and produces  a vector of action values $Q(s,\cdot;\theta)$, where $\theta$ represents the network's parameters. The loss function for DQN is defined as:
\begin{equation}
L(\theta) = \mathbb{E}\left[ \left(r + \gamma \max_{a'} Q(s',a';\theta^-) - Q(s,a;\theta) \right)^2 \right]
\end{equation}
where $\theta^-$ are the target network parameters, which are updated less frequently than the main network parameters $\theta$. The target network is used to stabilize the learning process by providing a fixed target for the Q-values

\section{Related Work}\label{sec2}

In quantum communication, maintaining high entanglement fidelity for remote connections has been identified as a critical factor for ensuring the reliability of quantum applications \cite{bib23}. Quantum repeaters face difficulties creating reliable entangled pairs due to background system noise, affecting the performance of quantum applications \cite{bib23}. This issue is particularly relevant in quantum cryptographic protocols like BB84, where the success of key distribution depemds on preserving entanglement fidelity above the threshold defined by quantum bit error rate (QBER) requirements \cite{bib24}.

The work in \cite{bib25} introduces a time-slotted model aimed at optimizing path assignments for pending requests in networks with limited resources. While the approach provides insights into traffic flow management and resource allocation under constrained conditions, it primarily addresses specific scenarios and lacks a broader application to more dynamic or diverse network conditions. The proposed scheduling mechanism offers a framework for resource prioritization, but its applicability to large-scale quantum networks with varying demands remains limited, highlighting the need for further exploration in this area.

Research in entanglement routing has largely concentrated on achieving high-throughput end-to-end connections \cite{bib17}, \cite{bib27}, \cite{bib28}. Although some efforts address network-wide throughput optimization for all source-destination pairs while considering constraints on entanglement quality \cite{bib2}, these approaches frequently neglect the practical limitations posed by the finite resources of quantum nodes. This oversight can significantly impact entanglement generation and swapping operations, underscoring gaps in addressing the resource constraints inherent in quantum networks.

Each source-destination pair in a quantum network may have distinct requirements, shaped by the needs of specific applications, leading to varying demands on entanglement distribution rates and time constraints. For instance, QKD necessitates continuous and stable entanglement. The understanding of these diverse application needs is essential for developing effective solutions that optimize network performance and meet the specific requirements of each use case.

Routing paths in quantum networks are determined by quantum entanglement, which is enabled through quantum repeaters and quantum swapping techniques \cite{bib29}. Optimization models are used to allocate qubits and repeaters to support various quantum entanglements \cite{bib30}, \cite{bib31}. However, although these aspects have been explored individually, there remains a significant gap in comprehensive research focused on designing routing networks tailored to the specific requirements of quantum applications.

Machine learning applications in quantum communication have been explored to a limited extent. Much of the current research focuses on optimizing entanglement durations and routing paths \cite{bib18}, \cite{bib20}, \cite{bib21}. While there is clear potential for machine learning techniques to improve quantum network routing efficiency \cite{bib38}, \cite{bib39}, notable gaps remain, such as the lack of consideration for reusing existing entanglements to boost overall network throughput, as seen in \cite{bib38}. Addressing these issues, machine learning has the potential to enhance the efficiency, reliability, and adaptability of quantum communication protocols.

Some studies, such as \cite{bib12}, overlook the degradation of fidelity during entanglement swapping. However, recent advancements in quantum repeaters have led to more robust and efficient communications \cite{bib40}. These advancements include the accommodation of additional qubits in quantum devices, enabling entanglement generation, purification, and the incorporation of quantum error correction techniques. Furthermore, by using multiple routing flows has been shown to improve performance \cite{bib41}. Today's internet structure accommodates many different applications in the same architecture. The quantum internet will also accommodate the   coexistence of diverse applications within a quantum network, and this will present interesting research opportunities. Addressing challenges in efficient resource allocation, routing, and prioritization will be critical for enabling future quantum applications, much like the role of TOS in ensuring quality and reliability in classical networks today.

\section{Quantum DQN Framework}\label{sec6}

The Quantum Deep Q-Network (QuDQN) framework combines quantum communication principles with a reinforcement learning model based on DQN. The design of this framework takes advantage of key quantum computing features, such as quantum entanglement, entanglement swapping, and quantum teleportation. At the same time, it uses the decision-making and learning capabilities of reinforcement learning agents. The core of the framework is a Neural Network (NN), which acts as a function approximator for Q-values. This allows the framework to process and represent quantum states, adapting the DQN algorithm to work in a quantum context where Q-values are estimated based on quantum information.

The QuDQN model incorporates key parameters for quantum networks, including network topology, qubit capacity at each node, and the set of requests (or demand) generated by quantum applications. It also considers factors like entangled bit (ebit) generation and swapping probabilities, as well as the fidelity thresholds needed for entangled pairs \cite{bib2}, \cite{bib10}, \cite{bib35}, \cite{bib40}. The adaptive routing agent in the framework is designed to handle entanglement distribution and task scheduling. It does this by managing entanglement routes and optimizing resource allocation across the quantum network. The goal is to create a flexible and reliable routing solution that learns from the quantum network environment, adapts to changes, and makes smart decisions to achieve the required fidelity for various quantum tasks.

\subsection{System Model}\label{subsec1}

In this work, we adopt the model proposed in \cite{bib39}. The system model examines a quantum network consisting of quantum repeaters connected through lossy links. The network's physical structure is represented by an undirected graph 
$G=(V,E)$, where $V$ represents the set of nodes, denoted as $V = \{ v_i \}_{i=1}^{N}$, and $E$ represents the set of edges, defined as $E = \{ e(i,j) \mid v_i, v_j \in V \}$. Each edge $e(i,j) \in E$ indicates the possibility of forming a quantum channel between nodes $v_i$ and $v_j$, enabling the establishment of quantum entanglement between their corresponding qubits. The distance between these nodes is within the entanglement threshold, set at approximately 1120 km \cite{bib45}, ensuring successful qubit entanglement.

\subsection{QuDQN Model}\label{subsec1}

At a given cycle $t$, the QuDQN model receives inputs, including the current network represented by $G^{(t)} = (V, E^{(t)})$, the number of qubits for each node $C^{(t)}$, the set of demands $D^{(t)}$, the ebit generation and swapping success probabilities $p_e$ and $q_v$, and the desired fidelity $F$, as shown in Figure 1. These inputs are used by the model to make decisions on resolving specific requests and taking appropriate actions. The model then uses a shortest-path component to efficiently route the selected requests. By incorporating these elements into the system model, we aim to optimize entanglement routing, resource allocation, and connection establishment in the quantum network, ultimately improving the overall performance and efficiency of quantum communication tasks.

\begin{figure}
\centering
\includegraphics [width=1\linewidth]{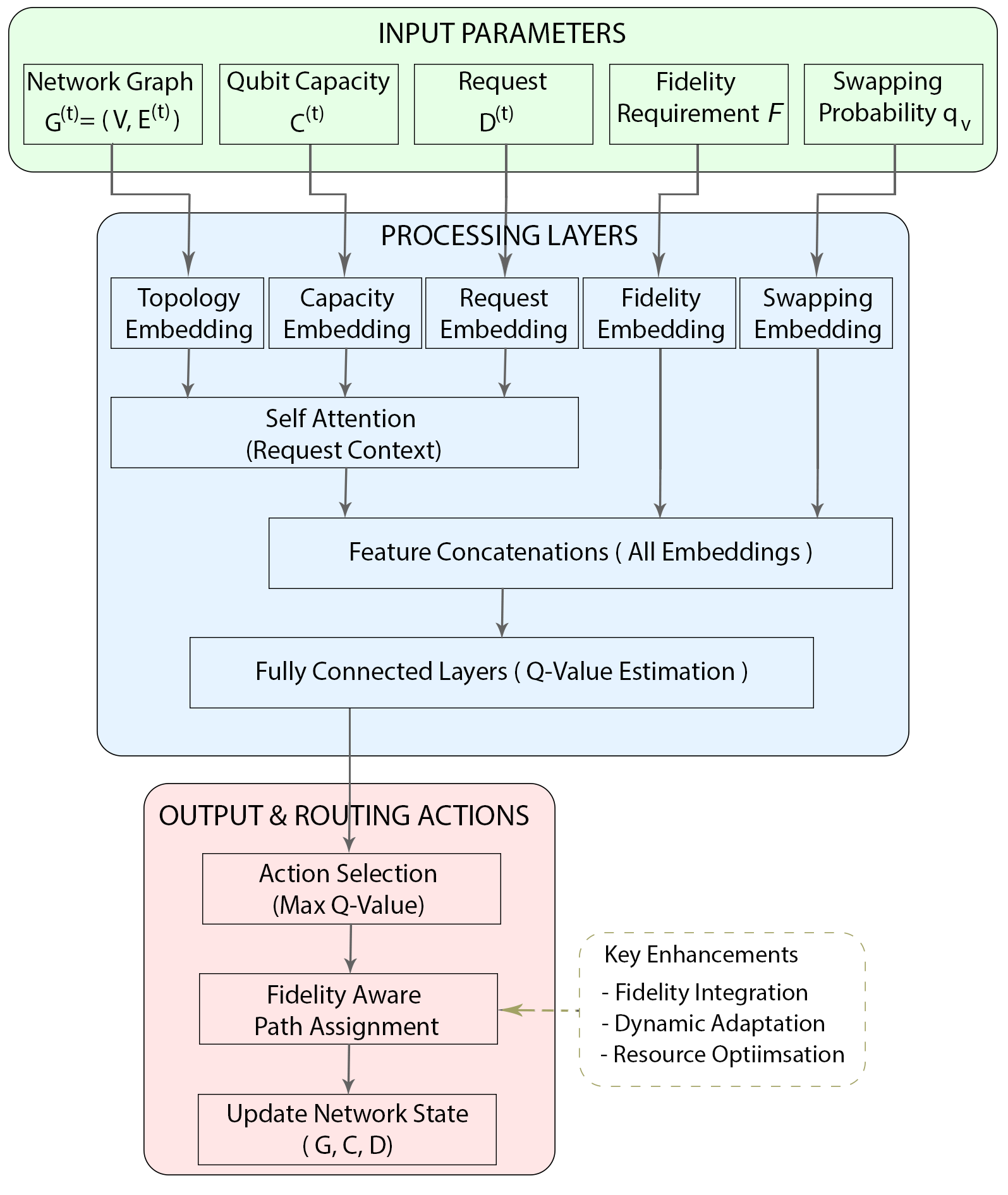}
\caption{Architecture of the QuDQN model, illustrating input parameters (network topology, qubit capacity, requests) and output actions (routing schedules, path assignments).}
\end{figure}

Each vertex \( v \in V \) is associated with a swapping success probability \( q_v \in (0, 1] \). Similarly, each communication channel has a success probability \( p_e \in (0, 1] \) for generating entanglement between nodes. In real-world implementations, the initial fidelity of a Bell pair is influenced by the underlying technology and can be modeled as a stochastic process with an expected value \( F \). This fidelity metric can be incorporated into the edge properties for all \( (i, j) \in E \) to enhance the network's characterization.

To address the fidelity requirements of specific quantum applications, we integrate success metrics for qubit entanglement and swapping into the model's input states. Additionally, we modify the reward function to account for critical factors such as fidelity and entanglement rates. With these enhancements, the QuDQN model's output action \( A^{(t)} \) is defined as follows: QuDQN generates actions that determine the routing schedule for requests in \( D \) and allocate their corresponding paths. These actions also consider \( E^{(t)} \), which includes success probabilities for qubit entanglement and swapping, as well as the fidelity requirements of the target applications. At each time slot \( t \) within an episode, QuDQN selects a request from the set of pending requests \( D^{(t)} \) and assigns a path connecting the source and destination nodes. This path allocation is guided by input states that incorporate fidelity and entanglement rate metrics.

We design a reward function (Eq. \ref{eq:11}) to align with our objectives. Specifically, the rewards obtained by the QuDQN model are directly tied to the number of successfully resolved requests, both in the current and future states of an episode. This reward structure encourages the routing agent to create schedules that maximize the number of fulfilled requests while penalizing schedules that result in fewer resolved requests. The reward function is carefully crafted to meet the core requirements of quantum applications, including fidelity and entanglement rate criteria.

\begin{widetext}
  \begin{equation}
    R^{(t)} = n^{(t)}_{r} \cdot \alpha + (|D| - n^{(t)}_{r}) \cdot \beta \cdot f 
    + \lambda \cdot R^{(t)} \cdot (1 - f) 
    + \gamma \cdot F^{(t)} \cdot q_v \cdot p_e
   \label{eq:11} 
  \end{equation}
\end{widetext}

Where $n^{(t)}_{r}$  represents the resolved requests after step $t$. The  reward term is given as $\alpha$, which emphasizes the importance of resolving requests.
$|D|$ are the total number of requests in the episode.
$\beta$ represents the penalty term, penalizing incomplete episodes. $f$ is a binary indicator that equals 1 when $t$ is the last step of an episode and 0 otherwise. The $\lambda \in[0,1]$ is the discount factor that controls the influence of future rewards on the current step. $R^{(t+1)}$ denotes the expected reward at the next time step. $\gamma$ is the  weighting factor for the fidelity and entanglement rates. $F^{(t)}$ the initial fidelity of a Bell pair, a stochastic process with expectation $F$. $q_v$  and $p_e$ are the swapping success probability associated with node $v$, the success probability for entanglement generation along the channel $e$ respectively. The reward function also punishes every wrong action made by the agent towards the end goal.

The QuDQN model selects the optimal path for a request based on the input states and the reward function. An equation for the optimal action \(A^{(t)}\) at a given step \(t\)
is :
\begin{equation}
A^{(t)} = \arg\max_{a \in \mathcal{A}} \Bigg[ Q\Big(G^{(t)}, C^{(t)}, D^{(t)}, p_e, q_v, F \Big) + R^{(t)} \Bigg]
\end{equation}

In this environment, $a \in \mathcal{A}$ denotes the action space for routing schedules. The Q-value function is represented by $Q(\cdot)$, which measures the expected cumulative reward after taking action $a$ at current states. In addition, $R^{(t)}$ represents the reward obtained at time step $t$, as defined in the proposed reward function.

\subsection{QuDQN Routing Approaches}
To address diverse operational scenarios and compare different routing strategies in quantum networks, we developed two baseline models under the QuDQN framework: QuDQN-Shortest and QuDQN-Random. These variations share the core architecture of QuDQN (Fig. 1) but uses distinct routing policies to evaluate different network conditions and providing flexibility for different deployment scenarios.

The QuDQN-Shortest strategy operates by selecting the shortest available path between source and destination nodes. For a network graph $G^{(t)} = (V, E^{(t)})$ and pending requests $D^{(t)}$, this method identifies the path $P_{\text{shortest}}^{(t)}$ that minimizes cumulative edge weights $w_e^{(t)}$, such as hop counts. The path is chosen as:

\begin{equation}
P_{\text{shortest}}^{(t)} = \arg\min_{P \in P(s,d)} \sum_{e \in P} w_e^{(t)},
\end{equation}

where $P(s,d)$ represents all possible paths between nodes $s$ and $d$. Requests are resolved greedily until qubit capacities $C^{(t)}$ are exhausted, leveraging Dijkstra’s algorithm for efficiency. While effective in small-scale or low-traffic networks, this approach neglects real-time resource constraints and fidelity requirements, leading to suboptimal performance in dynamic settings.

The QuDQN-Random strategy introduces stochasticity to explore routing diversity. At each time step $t$, it randomly selects a request $r \in D^{(t)}$ with uniform probability:

\begin{equation}
r_{\text{selected}}^{(t)} \sim \text{Uniform}(D^{(t)}),
\end{equation}

and routes it via the shortest available path $P_{\text{shortest}}^{(t)}$. This method prioritizes exploration over optimization, ignoring fidelity thresholds and resource limitations. While useful for evaluating network robustness, its lack of strategic prioritization often results in inefficient qubit utilization and unresolved high-priority requests.

Both baseline strategies use the core QuDQN architecture (Fig. 1) but replace the adaptive Q-value policy (Eq. 12) with static or stochastic rules. Their reward functions follow the general form of Eq. 11.  For QuDQN-Shortest, requests are resolved sequentially via precomputed shortest paths. For QuDQN-Random, requests are permuted randomly and routed only if resources permit. The output includes resolved requests $n_r^{(t)}$ and consumed qubits $Q_{\text{used}}^{(t)}$.

To illustrate the operational contrasts between routing strategies, three network configurations were analyzed (Figs. 2–4), each representing distinct routing methodologies under identical initial conditions. The network comprises 20 nodes arranged in a grid topology, with qubit capacities randomized between 4–6 units to emulate current quantum hardware limitations. A total of 6 requests were created for the three different network models to test the number of requests that will be fulfilled with the strict resource constraints.

\begin{figure}
\centering
\includegraphics [width=1\linewidth]{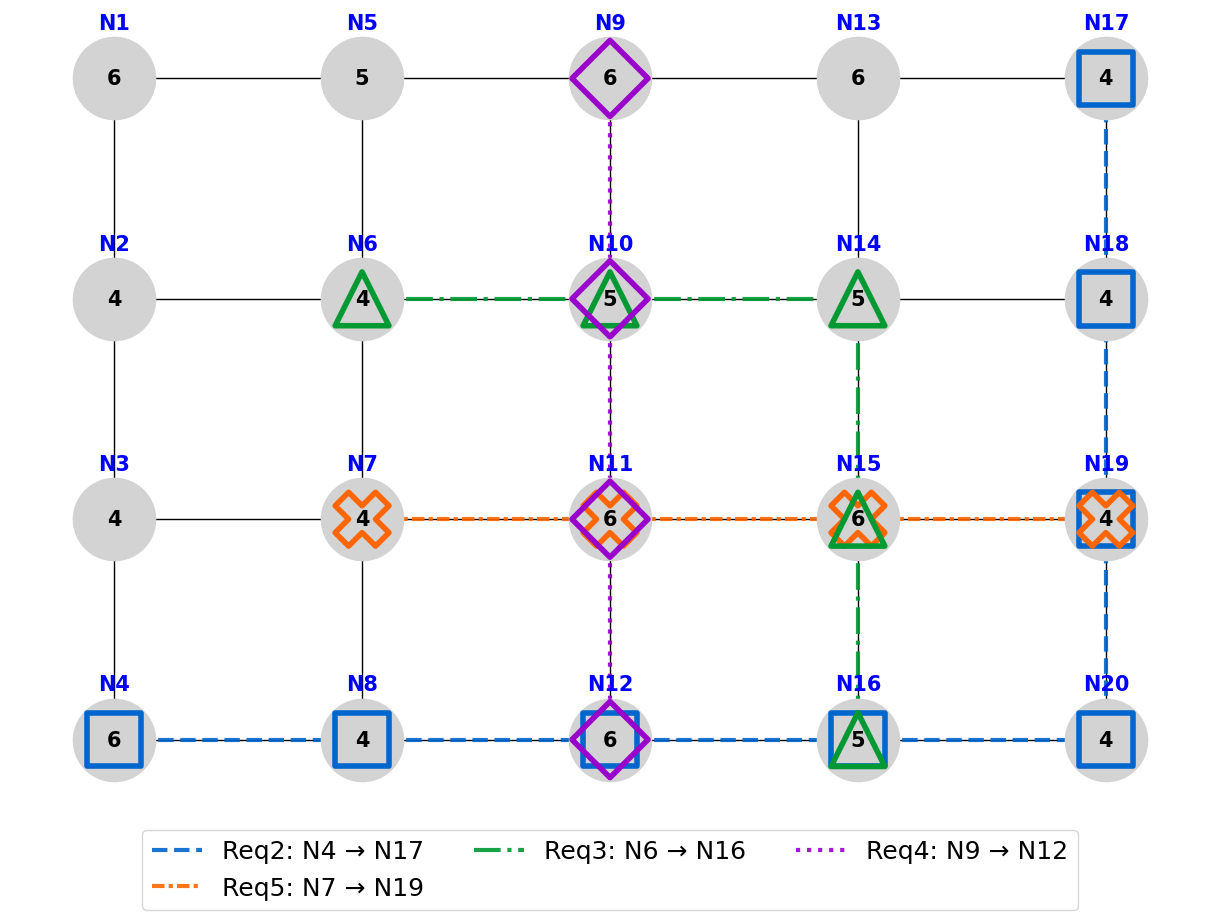}
\caption{Random routing (QuDQN-Random) results in uneven resource distribution and stranded capacity.}
\end{figure}
\begin{figure}[H]
\centering
\includegraphics [width=1\linewidth]{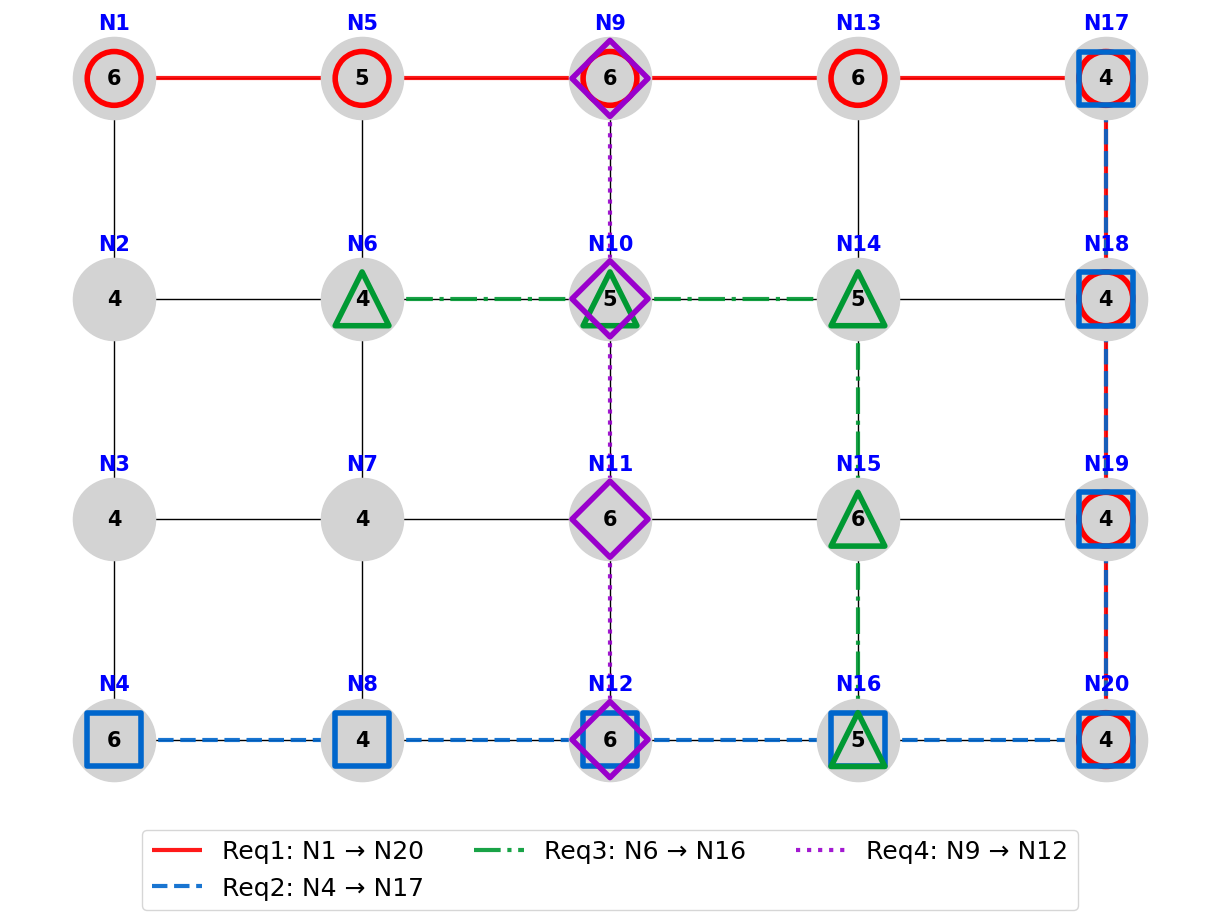}
\caption{Shortest-path routing (QuDQN-Shortest) depletes central nodes through repetitive path selection.}
\end{figure}
Figure 2 (QuDQN-Random) highlights the inefficiencies of stochastic routing, where arbitrary path selection leads to scattered resource exhaustion (e.g., N16: 3 qubits) despite underutilizing critical nodes. Figure 3 (QuDQN-Shortest) reveals the drawbacks of rigid shortest-path routing: repeated use of optimal paths rapidly depletes central node resources (N11: 6 qubits; N12: 6 qubits), mirroring patterns observed in classical network congestion. Figure 4 (QuDQN) demonstrates balanced resource utilization, where critical nodes retain significant residual capacity (N11: 6 qubits; N12: 6 qubits) even after routing multiple requests. This preservation stems from the adaptive path selection, which avoids overburdening nodes while maintaining efficient connectivity. 

The visualizations emphasize how adaptive routing (Fig. 4) mitigates resource exhaustion (Fig. 3) and fragmented allocation (Fig. 2). This is achieved by dynamically adjusting to real-time resource availability and request priorities, the QuDQN framework ensures sustainable network operation.

\begin{figure}
\centering
\includegraphics [width=1\linewidth]{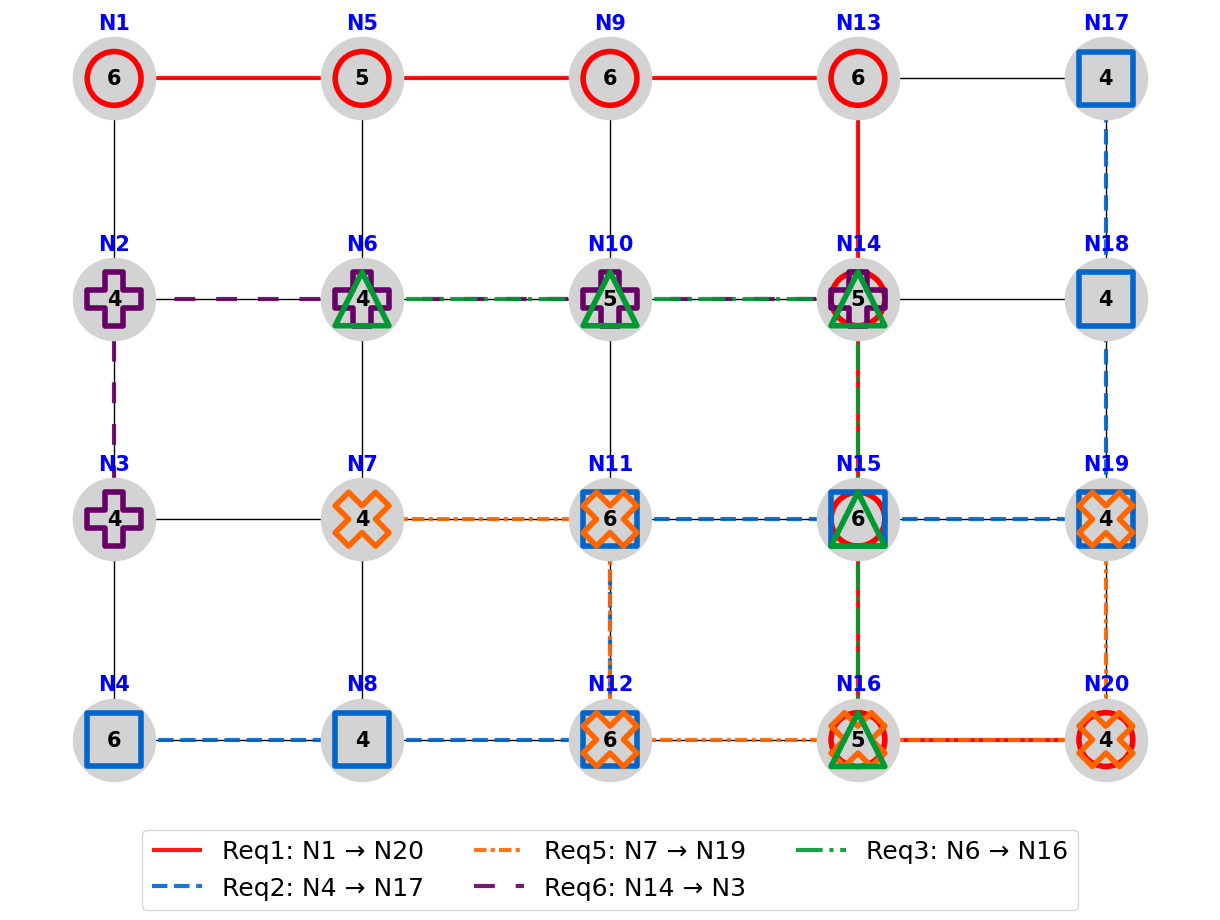}
\caption{Adaptive routing (QuDQN) preserves critical node resources (N11, N12) while fulfilling requests.}
\end{figure}

\subsection{QuDQN's Routing Efficiency Mechanisms}

The QuDQN model’s superior performance stems from three interconnected design principles. First, its adaptive reward function with fidelity prioritization (Eq. 11), explicitly rewards resolved requests, penalizes unresolved tasks, and prioritizes high-fidelity paths. This multi-objective formulation ensures alignment with quantum application requirements, such as QKD’s fidelity thresholds and time-sensitive task deadlines. The state further enhances adaptability by encoding real-time topology changes, qubit availability, and probabilistic success metrics, enabling context-aware routing.

Second, the Q-value-guided path selection uses a masked action policy to avoid redundant allocations. For each request batch, the agent ranks candidate actions by descending Q-values:
\[
a_t = \arg \max_{a \in \mathcal{A} \setminus \mathcal{A}_{\text{failed}}} Q(s_t, a; \theta)
\]  
where \(\mathcal{A}_{\text{failed}}\) denotes actions leading to invalid paths (e.g., insufficient qubits or failed swaps). Intermediate nodes deduct 2 qubits only if entanglement swapping succeeds, while end nodes deduct 1 qubit, minimizing resource waste. This ranking process, formalized as \(\text{sort}(Q) \rightarrow Q_{\text{sorted}}\), ensures that it explores the most promising paths first, maximizing efficiency and avoiding short-sighted decisions that can occur with simpler methods like always choosing the shortest path or randomly selecting routes. This approach allows the model to make smarter, long-term routing choices, balancing immediate gains with future resource availability. ensures efficient exploration of high-reward paths while avoiding myopic decisions inherent in greedy or random baselines.

Third, target network stabilization helps keep the learning process stable by periodically updating the target network's parameters. This is done at regular intervals, every 100 steps, given by
\[
\theta^{-} \leftarrow \theta \quad \text{at intervals} \quad t = kT \quad (k \in \mathbb{N}, T = 100),
\]
Here, \(\theta^{-}\) represents the parameters of the target network,  while \(\theta\)  represents the parameters of the main network. By updating the target network separately from the main network, this approach reduces errors caused by rapid changes during training. This is especially important for stabilizing the learning process in large-scale quantum networks. Together, these features allow QuDQN to effectively balance speed, accuracy, and resource usage. The advantages that are proven in Section 4 through comparisons with other leading methods.

\section{Performance Evaluation}\label{sec7}
The performance of the proposed quantum network is evaluated through simulations. The setup supports the generation, storage, exchange, and processing of quantum information through nodes interconnected via optical fiber links, as demonstrated in prior studies \cite{bib12, bib42}.  The routing protocol is tested across diverse real-world scenarios to assess its robustness and effectiveness under practical conditions. These simulations demonstrate the protocol’s ability to maintain reliable performance despite varying network demands, confirming its adaptability to different network designs. The results emphasize the protocol’s versatility, showcasing its potential for seamless integration into a wide range of quantum communication systems.

The QuDQN routing model is benchmarked against two simplified variants: the QuDQN-Shortest and QuDQN-Random models. Both variants are trained and constructed using the QuDQN framework but uses distinct routing protocols. The QuDQN-Shortest model prioritizes minimal path lengths, allocating routes using a shortest-path algorithm for all incoming requests. In contrast, the QuDQN-Random variant adopts a stochastic strategy, randomly selecting requests from the pending pool before routing them via the shortest available path.

Additionally, the QuDQN model is evaluated against state-of-the-art algorithms, including EFiRAP \cite{bib17} and ERDREG-PU \cite{bib26}. The EFiRAP method focuses on maximizing network throughput by optimizing entanglement link distribution across multiple source-destination pairs. ERDREG-PU, a progressive routing algorithm designed for quantum lattice networks, employs parallel path allocation and entanglement purification to maintain fidelity under finite edge capacity constraints.

A comprehensive analysis is conducted using four key metrics: request completion, network throughput, quantum channel utilization, and qubit utilization. Request completion quantifies the number of successfully resolved requests per episode, while network throughput measures the entanglement generation rate across the network. Quantum channel utilization reflects the proportion of actively used communication links, and qubit utilization tracks the number of qubits consumed during entanglement generation and swapping operations. Together, these metrics provide a rigorous technical assessment of the QuDQN model’s operational efficiency, scalability, and resource management capabilities, offering critical insights into its performance relative to existing approaches in quantum network deployments.

\subsection{Simulation Environment}\label{subsec1}

All simulations were implemented in Python using essential Python libraries. The quantum network was modeled and visualized with NetworkX, where nodes represent quantum devices (e.g., quantum repeaters) and edges correspond to quantum channels connecting these devices. The library’s built-in algorithms, such as those for computing shortest paths, were leveraged to route entanglement requests across the network. Numerical computations were handled using NumPy, and TensorFlow was used to construct the DQN architecture. Simulations were done on a workstation equipped with an 11th Gen Intel\textregistered{} Core\texttrademark{} i7-11850H CPU, 16 GB of RAM, and an NVIDIA RTX A3000 GPU to ensure sufficient computational capacity.

To evaluate the QuDQN model’s performance, the following parameters were defined: the entanglement generation and swapping success probabilities were fixed at 0.9. Key hyperparameters—$\alpha$, $\beta$, $\gamma$, and $\lambda$—were set to 0.2, $-1$, 0.9, and 0.9, respectively\cite{bib2, bib39}. The learning rate ($lr$) during training was configured to 0.1, and the mini-batch size was held constant at 512, consistent with the setup in \cite{bib39}. The fidelity $F_i$ of Bell pairs generated over quantum links followed a uniform distribution in the range $[0.70, 0.95]$ \cite{bib17}.

For experimental scenarios, the quantum nodes were assigned a variable number of quantum memory units, and quantum channel capacities were distributed between 26 and 35. The minimum required end-to-end fidelity $F$ for successful communication was set to 0.85, ensuring compliance with high-quality quantum communication standards. These configurations provided a robust testbed for assessing the QuDQN model’s ability to manage entanglement routing under diverse network conditions.

\subsection{Evaluation Results}\label{subsec1}

We first evaluate the viability of QuDQN against its variants, QuDQN-Random and QuDQN-Shortest. Tests were conducted over 500 simulation runs to compute the average number of successfully resolved requests and qubits utilized across all intervals. Results demonstrate that QuDQN consistently outperforms both variants in all scenarios. This superiority stems from QuDQN’s ability to incorporate alternative paths that account for both available qubit resources and path fidelity, enabling more informed and robust routing decisions.

\begin{figure}[h]
\centering  
\includegraphics[width=1\linewidth]{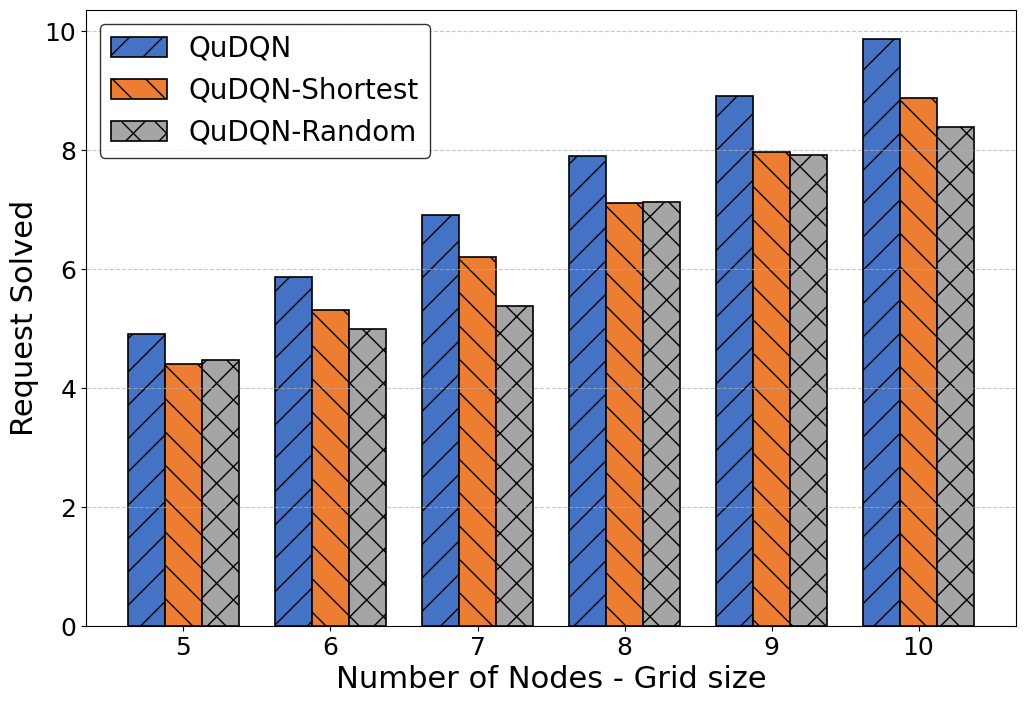}  
\caption{Comparative success rate of resolved requests as network size scales in grid-based topologies (5×5 to 10×10 nodes)}  
\label{fig2}  
\end{figure}

In the first scenario, we analyze how scaling the number of nodes and routing requests affects the performance of QuDQN, QuDQN-Random, and QuDQN-Shortest. This evaluation focuses on path allocation efficiency in a resource-constrained environment: each node is limited to four qubits, while grid sizes (ranging from \(5 \times 5\) to \(10 \times 10\)) and routing requests (increased incrementally from 5 to 10) are scaled. Figure~\ref{fig2} illustrates the average number of resolved requests across demand scenarios. QuDQN resolves 99\% of requests in all test cases, significantly outperforming QuDQN-Random (\(\sim\)80\%) and QuDQN-Shortest (\(\sim\)75\%). In the \(5 \times 5\) grid, QuDQN surpasses QuDQN-Shortest by 10.19\% and QuDQN-Random by 8.95\%. For the \(10 \times 10\) grid, the performance gap widens to 10.00\% against QuDQN-Shortest and 15.03\% against QuDQN-Random. This enhanced performance arises from QuDQN’s strategic pooling of requests, which evaluates all feasible paths prior to resource allocation. By holistically analyzing network conditions, QuDQN optimizes path selection based on real-time resource availability and fidelity constraints.  

\begin{figure}[h]
\centering  
\includegraphics[width=1\linewidth]{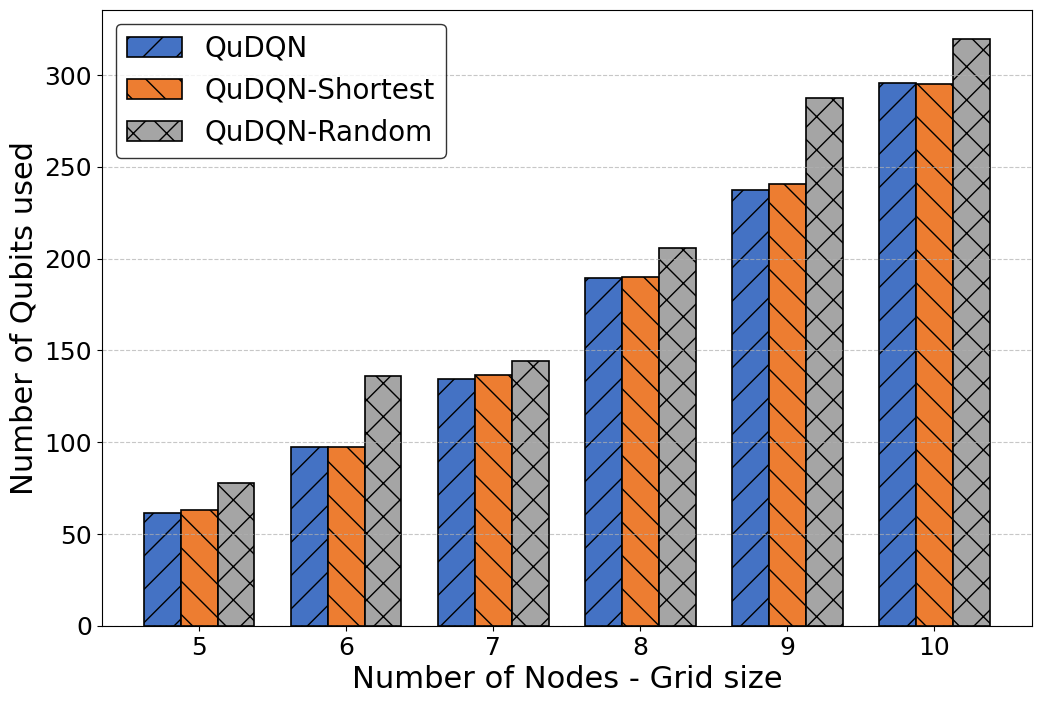}  
\caption{ Qubit utilization efficiency across varying network sizes and routing demands, highlighting resource conservation in QuDQN}  
\label{fig3}  
\end{figure}

Figure~\ref{fig3} illustrates the total qubit utilization across all tested models. QuDQN exhibits significantly lower qubit consumption compared to its variants, even as network size and request volume scale. Specifically, QuDQN uses 30\% fewer qubits than QuDQN-Random and 25\% fewer than QuDQN-Shortest.

In the \(5 \times 5\) node configuration, QuDQN demonstrates 2.46\% greater efficiency than QuDQN-Shortest and 26.16\% greater efficiency than QuDQN-Random. For the \(10 \times 10\) grid, QuDQN matches QuDQN-Shortest in efficiency but outperforms QuDQN-Random by 8.13\%. These results underscore QuDQN’s capability to operate efficiently in memory-constrained environments, a critical requirement for practical quantum networks.

We further evaluate QuDQN against the EFiRAP \cite{bib17} and ERDREG-PU \cite{bib26} algorithms. A robust quantum network must maximize throughput, optimize resource utilization, and handle high request volumes. To test this, we incrementally increased the number of source-destination pairs from 10 to 30 under fixed conditions: a \(7 \times 7\) grid, 20 qubits per node, and constant link capacities.  

\begin{figure} [h]
\centering  
\includegraphics[width=1\linewidth]{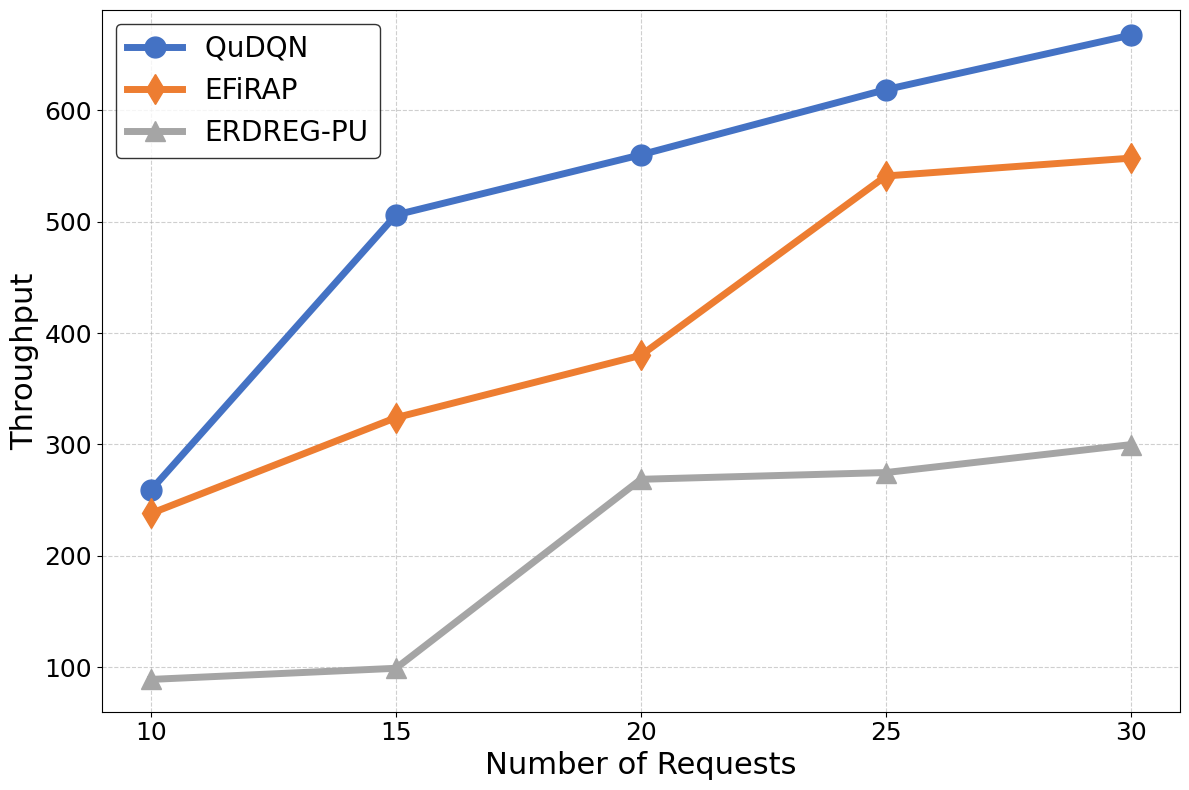}  
\caption{Throughput performance under escalating request loads (10 to 30 requests), demonstrating QuDQN’s superiority over EFiRAP and ERDREG-PU.}  
\label{fig4}  
\end{figure}

Figure~\ref{fig4} shows the throughput comparison. QuDQN resolves 99\% of requests across all demand levels (10–30 requests), surpassing both EFiRAP and ERDREG-PU. At 10 requests, QuDQN outperforms EFiRAP by 8.11\% and ERDREG-PU by 65.64\%. At 30 requests, the margin widens to 16.56\% against EFiRAP and 55.10\% against ERDREG-PU.

\begin{figure} [h]
\centering  
\includegraphics[width=1\linewidth]{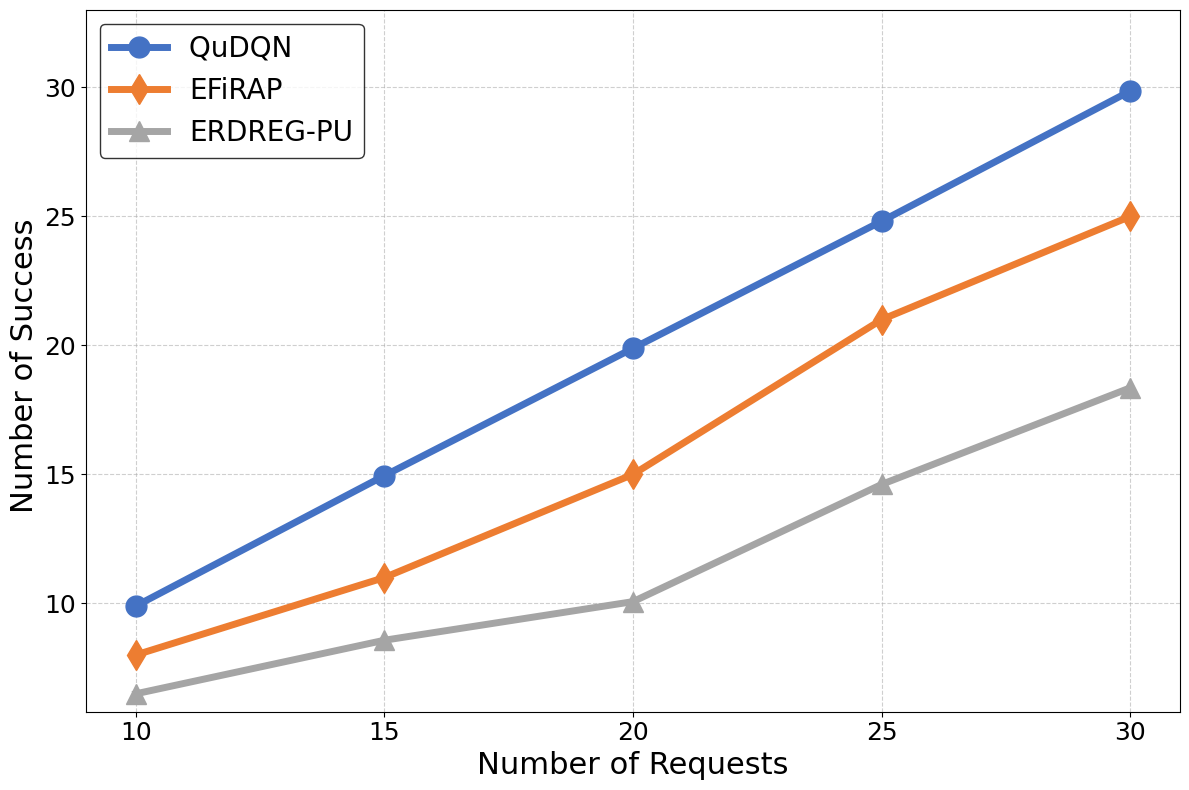}  
\caption{Average resolved requests per episode under increasing demand levels, showcasing QuDQN’s robustness in high-load scenarios.}  
\label{fig5}  
\end{figure}  
\begin{figure}[h]
\centering
\includegraphics[width=1\linewidth]{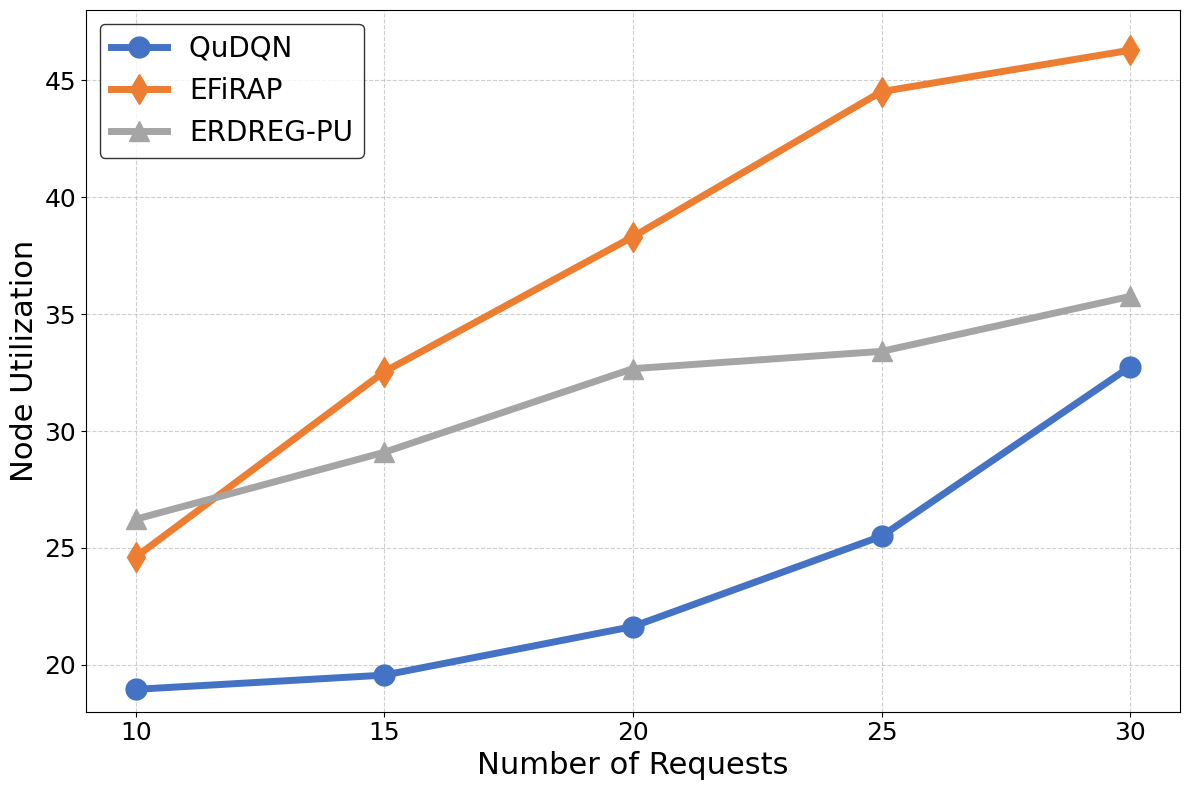}
\caption{Node qubit utilization across QuDQN, EFiRAP, and ERDREG-PU, demonstrating QuDQN's efficient memory management under varying request loads.}
\label{fig6}
\end{figure}

\begin{figure}[h]
\centering
\includegraphics[width=1\linewidth]{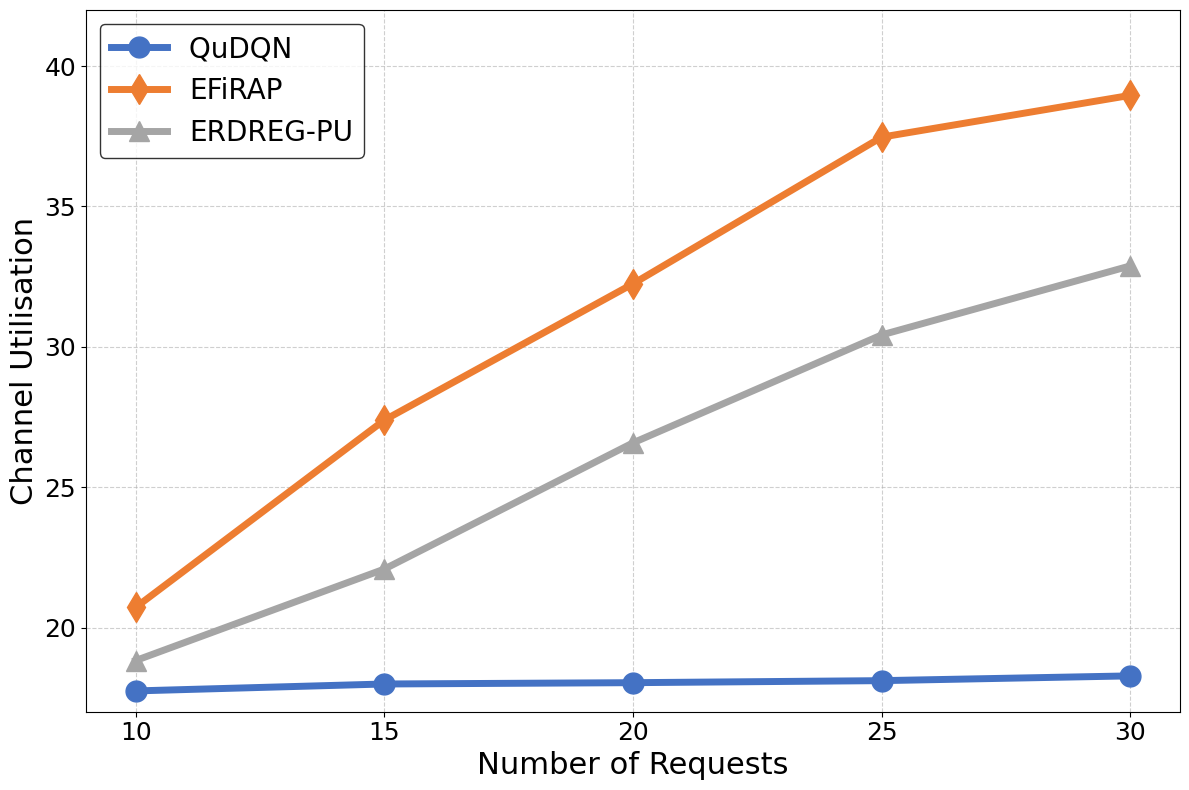}
\caption{ Link utilization efficiency, illustrating how reduced channel usage enables accommodation of future entanglement requests.}
\label{fig7}
\end{figure}
Figure~\ref{fig5} highlights QuDQN’s ability to manage higher request volumes. At 10 requests, QuDQN achieves a 23.73\% higher success rate than EFiRAP and 34.38\% higher than ERDREG-PU. At 30 requests, it maintains a 16.28\% advantage over EFiRAP and 38.54\% over ERDREG-PU. This efficiency stems from QuDQN’s dynamic routing strategy, which evaluates shared paths and allocates alternative routes when conflicts arise, ensuring robust resource management in complex quantum networks. 

To address the challenge of limited qubit capacity in current quantum systems, we analyze qubits utilization of nodes and communication links. Figures~\ref{fig6} and~\ref{fig7} compare the performance of QuDQN, EFiRAP, and ERDREG-PU in managing these critical resources. QuDQN consistently outperforms both benchmarks across all scenarios, demonstrating superior efficiency in resource-constrained environments.

As shown in Figure~\ref{fig6}, QuDQN optimizes quantum memory usage more effectively than competing models. At 10 requests, QuDQN consumes 18.96 qubits—29.85\% fewer than EFiRAP (27.03 qubits) and 38.34\% fewer than ERDREG-PU (30.74 qubits). This efficiency gap widens at higher loads: with 30 requests, QuDQN uses 32.75 qubits, representing a 41.28\% reduction compared to EFiRAP (55.78 qubits) and a 9.18\% improvement over ERDREG-PU (36.06 qubits).

Figure~\ref{fig7} illustrates QuDQN’s advantage in preserving channel capacity for future operations. At 10 requests, QuDQN uses 17.74 channels—16.79\% fewer than EFiRAP (21.33 channels) and 6.10\% fewer than ERDREG-PU (18.90 channels). This efficiency becomes more pronounced at scale: with 30 requests, QuDQN requires only 18.28 channels, outperforming EFiRAP by 113.17\% (39.02 channels) and ERDREG-PU by 79.91\% (90.71 channels). These results confirm QuDQN’s ability to maintain reserve capacity for subsequent entanglement tasks, a critical requirement for scalable quantum networks.

\section{Conclusion}\label{sec13}
The QuDQN architecture presents a groundbreaking framework for optimizing entanglement routing and resource allocation in quantum networks. By incorporating critical network parameters such as topology, qubit capacities, entanglement generation rates, and swapping success probabilities. The QuDQN dynamically addresses challenges posed by limited quantum resources and fidelity constraints. Central to its design is a reward function that balances resolved requests, fidelity requirements, and entanglement rates, enabling efficient routing decisions. This approach ensures robust performance even under stringent resource limitations, marking a significant advancement in quantum network management.  

Through extensive simulations, QuDQN exhibits significantly enhanced performance when compared to its simplified variants (QuDQN-Shortest and QuDQN-Random) as well as state-of-the-art algorithms (EFiRAP and ERDREG-PU). The results highlight QuDQN's ability to outperform these benchmarks, demonstrating its effectiveness in addressing the complexities of quantum network routing and resource allocation. The model achieves a 99\% request resolution rate across diverse network configurations, outperforming baseline models by 8–15\%. It also reduces qubit utilization by 25–30\% compared to alternative approaches, highlighting its efficiency in resource management. Furthermore, QuDQN enhances throughput and channel utilization, with improvements of 16–65\% over existing methods under high-demand scenarios. These results underscore QuDQN’s scalability in resource-constrained environments, such as nodes with limited qubit memory or networks with finite channel capacities. By prioritizing alternative paths and shared resource optimization, QuDQN offers a practical solution to the challenges of modern quantum communication systems. Future work will explore extensions to dynamic network topologies, hybrid classical-quantum routing protocols, and integration with real-world quantum hardware deployments.

\bibliographystyle{quantum}

\end{document}